\documentclass[epjST]{svjour}
\usepackage[colorlinks,citecolor=blue,urlcolor=blue,linkcolor=blue]{hyperref}
\usepackage{graphicx,colordvi}
\usepackage{booktabs,tabulary}
\usepackage{dsfont}
\usepackage{color}
\usepackage{epstopdf}  
\usepackage{fix-cm}

\usepackage{ulem}
\usepackage{graphics}

\newcommand{\be}{\begin{equation}}
\newcommand{\ee}{\end{equation}}
\newcommand{\wt}{\widetilde}
\newcommand{\beq}{\begin{equation}}
\newcommand{\eeq}{\end{equation}}
\newcommand{\bea}{\begin{eqnarray}}
\newcommand{\eea}{\end{eqnarray}}
\newcommand{\ba}{\begin{align}}
\newcommand{\ea}{\end{align}}
\newcommand{\bfig}{\begin{figure}}
\newcommand{\efig}{\end{figure}}

\newcommand{\wh}{\widehat}

\begin{document}
\title{Conformal mappings in perturbative QCD}
\author{Irinel Caprini\thanks{\email{caprini@theory.nipne.ro}}}
\institute{Horia Hulubei National Institute for Physics and Nuclear Engineering, P.O.B. MG-6, 077125 Bucharest-Magurele, Romania}
\abstract{We discuss  the method of conformal mappings applied to perturbative QCD. The approach is based on the  Borel-Laplace integral regulated with the principal value prescription and  the expansion of the Borel transform in powers of the variable which performs the conformal mapping of the cut Borel plane onto the unit disk.  We write down the expression of the conformal mapping  for the most general  location of the singularities of the Borel transform and review the properties of the corresponding expansions of the correlators.  Unlike the standard perturbative expansions, which are divergent, the modified expansions have a tamed behaviour at large orders and may even converge under some conditions. On the other hand, the expansion functions exhibit nonperturbative features similar to those of the expanded function. Using these properties, it was suggested recently that the expansions based on the conformal mapping of the Borel plane  may provide  an alternative to the standard OPE. We briefly review the arguments in favour of this conjecture and discuss the application of the method to the Adler function for masseles quarks and the static quark self-energy  calculated in lattice QCD.
} 
\maketitle
\section{Introduction}
\label{sec:intro}

It is known that perturbation theory leads to divergent series for many quantities in quantum mechanics (QM) and  in renormalized quantum field theory (QFT). This fact was  noticed for the first time for QED by Dyson \cite{Dyson:1952tj}, who proposed to interpret the divergent series as asymptotic to  the exact function. This assumption implies a profound change in the interpretation of perturbation theory: while in the case of a  convergent series the knowledge of all the perturbative coefficients  determines uniquely the expanded function, there are infinitely many functions having the same  asymptotic expansion. The problem is to choose the best approximant, which incorporates as much as possible of the known properties of the exact function.

The divergent pattern of perturbation theory in  QFT indicates a singular behaviour of the expanded Green functions at the origin of the  coupling  plane. For QED,  Dyson showed by a simple argument that the physical observables cannot be analytic at $\alpha=0$  \cite{Dyson:1952tj}. For QCD, a complicated singularity structure  of the correlators at $\alpha_s=0$ was  shown to follow  from unitarity, analyticity and renormalization-group invariance \cite{tHooft}. Alternatively,  particular classes of Feynman diagrams indicate a factorial growth of the expansion coefficients in both QED \cite{Lautrup:1977hs},\cite{Broadhurst:1992si} and QCD \cite{Beneke:1994qe},\cite{Beneke:1992ch},\cite{Beneke:1998ui}.  Recently, evidence for a factorial increase of the perturbative coefficients in QCD was provided also by lattice calculations \cite{Bauer:2011ws}.

The large-order increase of the expansion coefficients of a function are encoded in the singularities of its Borel transform in the Borel complex plane.  For QCD, the dominant singularities are the infrared (IR) renormalons, which are produced by the low momenta in the Feynman diagrams, and the ultraviolet (UV) renormalons, related to high momenta in the Feynman diagrams. In particular, the IR renormalons are situated on the integration range of the Laplace-Borel integral by which the original function is recovered from its Borel transform in the Borel summation method. Therefore this integral is not well defined, which in mathematical 
terms means that the series is not Borel summable. In physical terms, the  ambiguities of perturbation theory related to the IR renormalons  suggest  that this theory is not complete and must be supplemented by  nonperturbative terms in order to recover the exact function \cite{Beneke:1998ui},\cite{Mueller1985},\cite{Mueller:1993pa}. According to the general view, these terms are identified with the power corrections in the Operator Product Expansion (OPE) of the Green functions \cite{Shifman:1978bx}. 

As remarked recently, the power corrections have a mathematical interpretation  in the so-called hyperasymptotic theory (see \cite{BerryHowls},\cite{Howls},\cite{Dorigoni:2014hea} and references therein), being a first piece of a transseries, i.e.  a sequence of truncated series, each of them  exponentially small in the expansion parameter of the previous one,  which allow to recover the expanded function from its asymptotic divergent expansion. From the point of view of analyticity, the essential feature of the power corrections is that they exhibit a singular behaviour at the origin of the coupling plane.  On the other hand, starting from the divergent pattern of the perturbative expansions in QCD, one can construct modified expansions that incorporate in an intrinsic way, through the expansion functions, the singular behaviour at zero coupling. This can be achieved  by reordering the perturbative series by means  of a suitable conformal mapping.

The method of conformal mappings  is known in mathematics as a technique  for ``series acceleration'', i.e.  for increasing the  rate of convergence  of power series. 
  By expanding a function in powers of the variable that maps its analyticity domain onto a  disk, the new series converges in a larger region,  well beyond the convergence domain of the original expansion, and  has an increased  asymptotic convergence rate at points lying inside this domain.
The method has been applied  a long time ago  in particle physics to the expansions of scattering amplitudes  in powers of various kinematical variables  \cite{CiFi},\cite{Frazer}. 
More recently, applications of the conformal mappings to the perturbative expansions in QFT have been considered in Refs. \cite{Seznec:1979ev}-\cite{Caprini:2020lff}. In QCD, the use of a conformal mapping of the Borel plane for  suppressing the power corrections related to the large momenta in Feynman diagrams was suggested for the first time in \cite{Mueller:1993pa} and was applied to the Adler function in  \cite{Altarelli:1994vz}. The optimal conformal mapping, which achieves the analytic continuation  in the whole Borel plane,  was then found in  \cite{Caprini:1998wg},  The properties of the perturbative expansions of the Adler function improved by means of this mapping have been investigated in \cite{Caprini:2000js},\cite{Caprini:2001mn}, and applications of the method to $\tau$ hadronic decays have been discussed in \cite{Cvetic:2001sn}-\cite{Caprini:2020lff}.

 In the present paper we review the application of the method of conformal mappings to perturbative QCD, emphasizing the fact that it  provides a systematic procedure for recapturing nonperturbative features of the QCD correlators. The outline of the paper is as follows: in the next section  we write down the expression of the optimal conformal mapping  for a general location of the leading IR and UV renormalons in the Borel plane. We define also the corresponding new expansions of the QCD correlators, showing  how to incorporate the  nature of the leading singularities of the Borel transform, when this information is available. In Sect. \ref{sec:prop} we summarize the properties of the new expansions. In Sect. \ref{sec:OPE} we  review the arguments presented in \cite{Caprini:2020lff} in favour of the conjecture that the expansions based on conformal mappings might be an alternative to the standard OPE. In Sect. \ref{sec:applic} we briefly discuss two applications: the Adler function for massless quarks and the self energy of a static quark source.  Finally, Sect. \ref{sec:conc} contains our conclusions.

\section{Conformal mappings of the Borel plane}\label{sec:conf}
  We consider a generic observable   $F(a)$, expressed in QCD perturbation theory by the  expansion  
\beq\label{eq:F}
F(a) =\sum\limits_{n=0}^\infty F_n a^{n+1},
\eeq
in powers of the renormalized strong coupling $a=a(\mu^2) \equiv \alpha_s(\mu^2)/\pi$, defined in a certain renormalization scheme  at the renormalization scale $\mu$. The dependence of $a(\mu^2)$ on the renormalization scale is governed by the renormalization-group equation
 \begin{equation}\label{eq:rge}
 -\mu^2\frac {d a(\mu^2)}
{d\mu^2}= \beta(a)=\sum_{n\ge 0}
\beta_n [a(\mu^2)]^{n+2}. \end{equation}
We recall that the first two coefficients of the above expansion  are renormalization-scheme independent and are expressed in terms of the number $n_f$ of active flavours as 
\beq\label{eq:beta01}
\beta_0=\frac{1}{4}\,(11-\frac{2}{3} n_f),\quad\quad \beta_1=\frac{1}{16}\,(102-\frac{38}{3} n_f).
\eeq

For many QCD observables, the coefficients $F_n$ exhibit a factorial increase $F_n\sim n!$   at high $n$. Additional factors specifying the  large-order behaviour of $F_n$ are known in some cases \cite{Beneke:1998ui}. It follows that the series (\ref{eq:F})  has zero radius of convergence and can be interpreted only as an asymptotic expansion to $F(a)$ for $a\to 0$. The divergent Taylor expansion indicates also the fact that the  function $F(a)$ is singular at the origin $a=0$ of the coupling plane. As mentioned above, for some observables, like the Adler function in massless QCD, this property was found by independent arguments based on renormalization-group invariance, unitarity and analyticity in the momentum plane \cite{tHooft}.

As will be explained below,  the method of series acceleration by conformal mappings can be applied only if  the expanded function is analytic in a region around the expansion point.  Therefore,  the method cannot be used in QFT for the standard perturbative series in powers of the coupling, since the Green functions are singular at the origin of the coupling plane\footnote{To circumvent this difficulty, an ''order-dependent'' conformal mapping of the coupling complex plane was defined  in  \cite{Seznec:1979ev},\cite{ZinnJustin:2010ng}, by assuming that the singularity is  shifted away from the origin at each finite perturbative order, and tends to the origin only in the limit of an infinite number of terms.}.  However, the conditions of applicability are satisfied by the Borel transforms of the correlators. Starting from the expansion (\ref{eq:F}), we define the Borel transform $B_F(u)$ by the power series
\be\label{eq:B}
 B_F(u)= \sum_{n=0}^\infty  b_n\, u^n,
\ee
where the coefficients $b_n$ are  related to the perturbative coefficients $F_{n}$ by 
\be\label{eq:bn}
 b_n= \frac{F_{n}}{\beta_0^n \,n!}.
\ee
Here $\beta_0$ is the first coefficient of the $\beta$ function, defined in (\ref{eq:beta01}).

Using the definition (\ref{eq:B}), one can check that the function $F(a)$ defined by the expansion (\ref{eq:F}) is recovered formally from the Borel transform by the Laplace-Borel integral representation 
\be\label{eq:Laplace}
F(a)=\frac{1}{\beta_0} \,\int\limits_0^\infty  
\exp{\left(\frac{-u}{\beta_0 a}\right)} \,  B_F(u)\, d u\,.
\ee

\begin{figure}
\includegraphics[width=5cm]{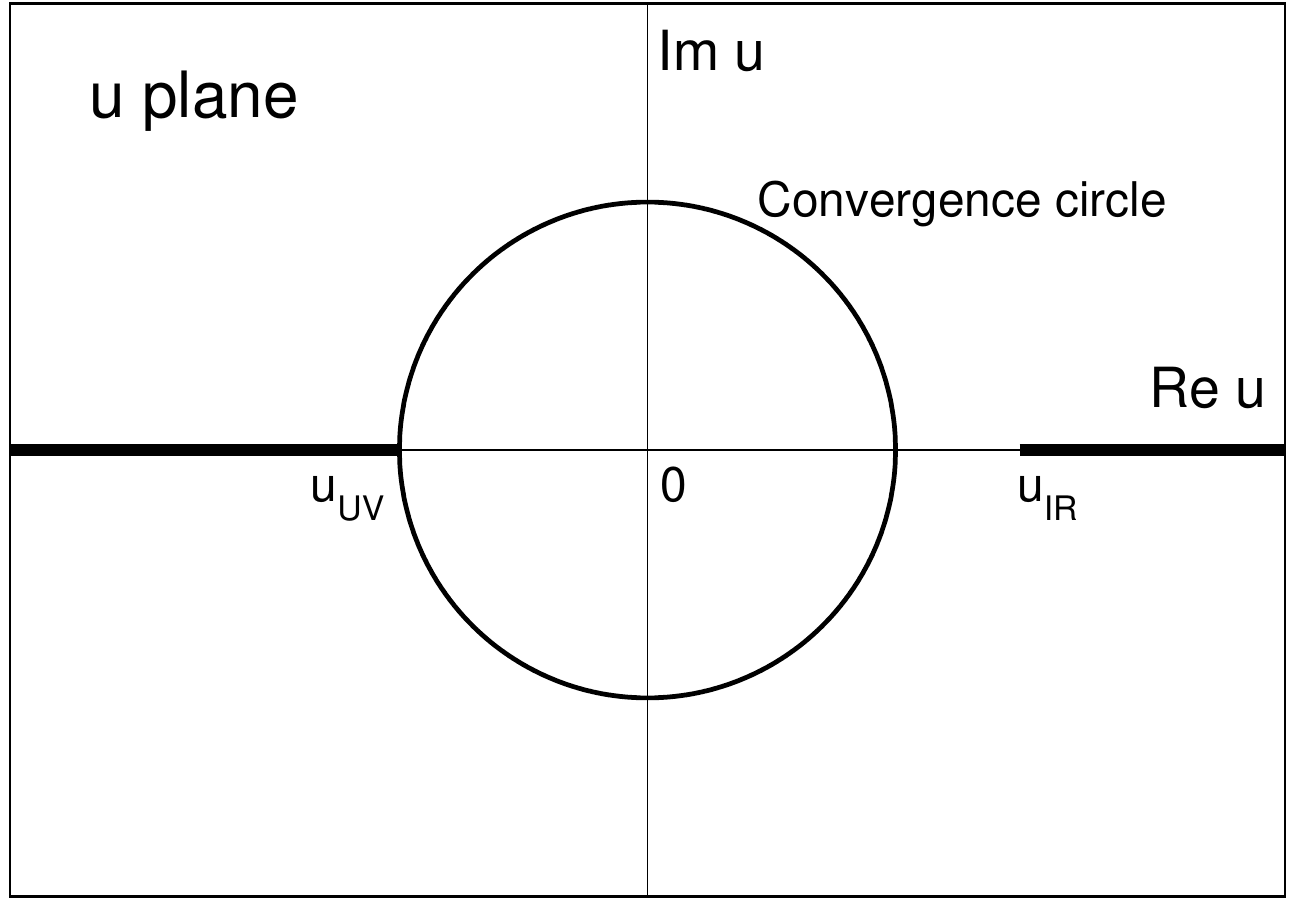}\hspace{1.5cm}\includegraphics[width=5cm]{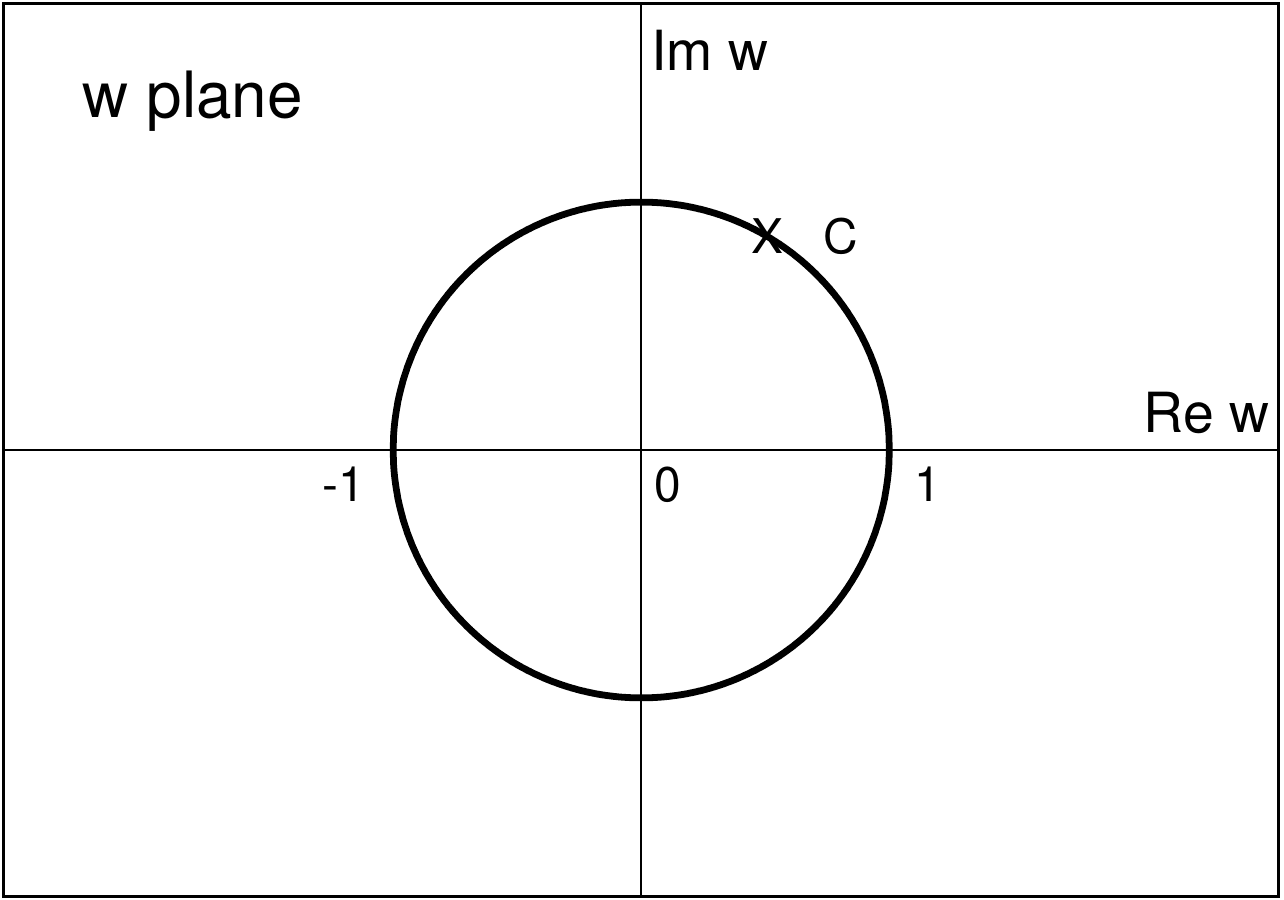}
\caption{Left: Borel plane. The series (\ref{eq:B}) converges inside the circle $|u|=|u_{UV}|$. Right: $w$ plane. All the UV and IR renormalons are situated on the circle $|w|=1$. $C$ is the image of the point at infinity.
\label{fig:Borel}}
\end{figure}

Since the coefficients $b_n$ defined in (\ref{eq:bn}) have a suppressed increase at large $n$, one expects the Taylor series (\ref{eq:B}) to be convergent in a region around the point $u=0$. Indeed, the singularities of the Borel transform in the complex $u$ plane, which encode the large-order increase of the coefficients $F_n$  of the perturbation series (\ref{eq:F}), are situated at a finite distance from the origin.  In the general case,  $B_F(u)$ has singularities at real values of $u$ on the semiaxes $u\ge u_{IR}>0$ (IR renormalons and instantons), and $u\le u_{UV}<0$ (UV renormalons).  In the large-$\beta_0$ limit  the singularities are poles, but beyond this limit they are branch points, requiring the introduction of two  cuts along the lines $u\geq u_{IR}$ and $u\leq u_{UV}$ (see  left panel of Fig. \ref{fig:Borel} for a typical Borel plane). The cuts along the real axis are assumed in general to be the only singularities in the Borel plane \cite{Mueller1985}. 

The singularities of $B_F(u)$ restrict the convergence domain of the power expansion (\ref{eq:B}) in the complex $u$ plane: this series converges only inside  the circle which passes through the singularity closest to the origin  $u=0$  (in the left panel of Fig. \ref{fig:Borel}, we show this circle, assuming the nearest singularity to be the first UV renormalon). 

Note that the integration range in (\ref{eq:Laplace}) extends for $u \ge |u_{UV}|$, beyond the convergence domain, this being actually the reason of the fact that the expansion (\ref{eq:F}) is divergent. 
Moreover, due to the singularities of $ B_F(u)$ for $u\ge u_{IR}$, the  integral (\ref{eq:Laplace}) is not defined and requires a regularization. As shown in \cite{Caprini:1999ma}, the principal value (PV) prescription, where the integral (\ref{eq:Laplace}) is defined as the semisum of the integrals along two lines parallel to the real axis $u\ge 0$, slightly above and below it, is convenient since it preserves to a large extent the analytic properties of the exact correlator in the  complex plane (in particular, it satisfies the requirement of giving a real result for real values of the coupling). Therefore, we shall adopt this prescription in what 
 follows. 

As mentioned in the Introduction, the domain of convergence of a power series in the complex plane can be enlarged and the convergence rate can be increased by expanding the function in powers of the variable which achieves the conformal mapping of the original plane  (or a part of it) onto a disk (which is the natural convergence domain of power series).   An important result, proved a long time ago \cite{CiFi}, is that the best asymptotic convergence rate is obtained by mapping  the entire holomorphy domain of the expanded function  onto the unit disk.

The key element in the proof presented in  \cite{CiFi} is the remark that the asymptotic convergence rate of a power series at a point in the complex plane is governed by the quotient $r/R$, where $r$ is the distance of the point from the origin and $R$ the convergence  radius.  Therefore, as argued in \cite{CiFi}, one must compare the magnitudes of the  ratio  $r/R$  for a certain point in different complex planes, corresponding to different
conformal mappings of the original plane. It turns out that when the whole analyticity domain  of the
function is mapped on a disk, the ratio $r/R$ reaches its minimum value (for a detailed proof see \cite{Caprini:2011ya}). This defines an ``optimal conformal mapping'', which leads to the best  asymptotic convergence rate of the corresponding expansion.

In the present case, we shall denote by $\tilde w(u)$ the optimal variable which performs the conformal mapping of the doubly-cut Borel plane  shown in the left panel of Fig. \ref{fig:Borel} onto the unit disk.
We recall that the expression of $\tilde w(u)$ was written down for the first time in \cite{Caprini:1998wg}, in the particular case $u_{UV}=-1$ and $u_{IR}=2$. The expression given in \cite{Caprini:1998wg} can be easily generalized to the arbitrary positions of leading singularities, shown in Fig. \ref{fig:Borel}: the optimal conformal mapping  is achieved by the function 
\begin{equation}\label{eq:w}
\tilde w(u)=\frac{\sqrt{1-u/u_{UV}}-\sqrt{1-u/u_{IR}}}{\sqrt{1-u/u_{UV}}+\sqrt{1-u/u_{IR}}},
\end{equation}
whose inverse reads
\beq\label{eq:uw}
\tilde u(w)= \frac{4 u_{UV} u_{IR} w}{u_{UV} - u_{IR} + 2 u_{UV} w + 2 u_{IR} w + u_{UV} w^2 - u_{IR} w^2}.
 \end{equation}

The function $\tilde w(u)$  maps the complex  $u$ plane cut along the real axis for $u\ge u_{IR}$ and $u\le u_{UV}$ onto the interior
of the circle $\vert w\vert\, =\, 1$ in the complex plane $w\equiv \tilde w(u)$,  such that  the origin $u=0$ of the $u$ plane
corresponds to the origin $w=0$ of the $w$ plane, and the upper (lower) edges of the cuts are mapped onto the upper
(lower) semicircles in the  $w$ plane (see the right panel of Fig. \ref{fig:Borel}, where we denoted by $C$  the image of the points $|u|\to \infty$ in the upper half plane). 
By the  mapping (\ref{eq:w}), all the  UV and IR  renormalons are placed on the boundary of the unit disk in the $w$  plane,  at equal distance from the origin. 

We finally note that if either UV or IR renormalons do not exist and the corresponding cut is absent, the  optimal variable performs the conformal mapping of a singly-cut Borel plane onto the interior of the unit circle. One can easily check that this mapping is obtained formally from (\ref{eq:w}), by letting the beginning of the missing cut to go to infinity. For instance, if there are no ultraviolet renormalons, the optimal mapping is given by 
\begin{equation}\label{eq:wIR}
\tilde w(u)=\frac{1-\sqrt{1-u/u_{IR}}}{1+\sqrt{1-u/u_{IR}}},
\end{equation}
and the inverse (\ref{eq:uw}) becomes
\beq\label{eq:uwIR}
\tilde u(w)= \frac{4  u_{IR} w}{(1- w)^2}.
 \end{equation}

 We consider now the expansion of $B_F(u)$ in powers of the variable $w\equiv \tilde w(u)$: 
\be\label{eq:Bw}
B_F(u)=\sum_{n= 0}^{\infty} c_n \,w^n.
\ee
Using Eqs. (\ref{eq:B}) and  (\ref{eq:w}), the coefficients $c_{n}$  are expressed in a straightforward way in terms of the coefficients $b_{k}$ with
$k\leq n$.   We emphasize that by expanding $B_F(u)$ according to (\ref{eq:Bw}), one makes full use of its
holomorphy domain, because the known part of it (the first Riemann sheet) is
mapped onto the convergence  disk.  Therefore, the series (\ref{eq:Bw}) converges in the whole $u$ complex plane up to the cuts, i.e. in a much larger domain than the original series (\ref{eq:B}).  Moreover, from the results mentioned above it follows that the expansion (\ref{eq:Bw}) has the best asymptotic convergence rate compared to other expansions, in powers of variables which map only a part of the  holomorphy domain onto the unit disk.

By inserting the  expansion (\ref{eq:Bw}) in the Borel-Laplace integral (\ref{eq:Laplace}), we are led  to a new perturbative series for the observable $F(a)$, of the form \cite{Caprini:1998wg},\cite{Caprini:2000js},\cite{Caprini:2001mn}
\begin{equation}
F(a)= \sum_{n=0}^\infty c_{n} {\cal W}_{n}(a), 
\label{eq:cW}
\end{equation}
where the functions ${\cal W}_n(a)$ are defined as 
\begin{equation}\label{eq:Wn}
{\cal W}_{n}(a)=\frac{1}{\beta_0} {\rm PV}\int\limits_0^\infty\, e^{-u/(\beta_0 a)}\, (\tilde w(u))^n \,du.
\end{equation}
Since we adopted the Principal Value prescription for regularizing the Borel-Laplace integral (\ref{eq:Laplace}) which gives the correlator $F(a)$, we use the same prescription for the expansion functions ${\cal W}_n(a)$.

We note that the expansion (\ref{eq:cW}) is obtained formally from  Eqs. (\ref{eq:Laplace}) and (\ref{eq:Bw}) by changing the order of integration and summation.  This procedure is trivially allowed when the series  (\ref{eq:Bw}) is truncated at  any finite order. For an infinite number of terms, however, the new expansion (\ref{eq:cW}) represents a nontrivial step out of perturbation theory, replacing the perturbative powers $a^n$ by the expansion functions ${\cal W}_{n}(a)$.

The expansion  (\ref{eq:cW}) can be further improved by exploiting  the available information on the behaviour of $B_F(u)$  in the Borel plane near the first renormalons located at $u=u_{UV}$ and $u=u_{IR}$. Near these  branch points, the generic behaviour of $B_F$ is
\begin{equation}\label{eq:rgamma}
 B_F(u) \sim \frac{r_{UV}}{(1-u/u_{UV})^{\gamma_{UV}}},  \quad  \quad  B_F(u)  \sim \frac{r_{IR}}{(1-u/u_{IR})^{\gamma_{IR}}}, 
\end{equation}
for $u\sim u_{UV}$ and $u\sim u_{IR}$, respectively, where the exponents
$\gamma_{UV}>0$ and $\gamma_{IR}>0$  have in general known expressions, involving the first coefficients (\ref{eq:beta01})  of the $\beta$ function. The residues $r_{UV}$ and $r_{IR}$ are also known in some cases.

The nature of the leading singularities can be implemented by considering the product $S(u) B_F(u)$, where  $S(u)$ is a ``softening factor''  which compensates the singular behaviour of $F(u)$ at $u=u_{UV}$ and $u=u_{IR}$.   The product remains finite at $u=u_{UV}$ and $u=u_{IR}$, but in general will still have branch points at these points.  Therefore, although the singularities are milder, the optimal variable for the expansion of the product is still the conformal mapping (\ref{eq:w}), which depends on the position of the first branch points.  Using this remark, we define the expansion
\be\label{eq:Bw1}
 B_F(u)=\frac{1}{S(u)} \sum_{n=0}^{\infty} {\widetilde c}_n\, w^n,
\ee  
where the coefficients $ {\widetilde c}_n$ are determined uniquely in terms of the coefficients $b_k$ with $k\leq n$ by an interative procedure.

By inserting (\ref{eq:Bw1}) in the Borel-Laplace integral (\ref{eq:Laplace}), and changing the order of integration and summation as discussed above,  we  define the expansion
\be\label{eq:cWtilde}
F(a)=\sum\limits_{n=0}^\infty {\widetilde c}_n {\widetilde {\cal W}}_n(a), 
\ee
where the expansion functions are
\be\label{eq:Wntilde}
{\widetilde {\cal W}}_n(a)=\frac{1}{\beta_0}{\rm PV} \int\limits_0^\infty e^{-u/(\beta_0 a)}\frac{(\tilde w(u))^n}{S(u)}du.
\ee

As emphasized in  \cite{Caprini:2009vf},\cite{Caprini:2011ya}, while the optimal conformal mapping (\ref{eq:w}) is unique,  the softening factor $S(u)$ is not. The problem was investigated in detail in \cite{Caprini:2011ya}. For instance, one can take $S(u)$ as the simple product
\be\label{eq:Su}
S(u)=(1-u/u_{UV})^{\gamma_{UV}}(1-u/u_{IR})^{\gamma_{IR}}, 
\ee
or a suitable expression in terms of the variable $w$: 
\be\label{eq:Sw}
S(u)=(1+\tilde w(u))^{2\gamma_{UV}}(1-\tilde w(u))^{2\gamma_{IR}},
\ee
which has the same behaviour near $u_{UV}$ and $u_{IR}$.
\section{Properties of the expansions based on conformal mappings}\label{sec:prop}
The properties of the expansions based on the conformal mapping of the Borel plane have been discussed in detail in Refs. \cite{Caprini:2000js}, \cite{Caprini:2001mn}, \cite{Caprini:2011ya} for the particular case of the Adler function in massless QCD. These properties are valid also in the more general case discussed in this paper, of arbitrary positions of the dominant singularities in the Borel plane. In this section we shall briefly review these properties.  

We note first that the expansion functions (\ref{eq:Wn}) and  (\ref{eq:Wntilde}) are no longer powers of the coupling, as in the standard perturbation theory, and exhibit a nontrivial dependence on $a$. To emphasize this fact, in Refs. \cite{Abbas:2013usa},\cite{Caprini:2020lff}  the new expansions  (\ref{eq:cW}) and  (\ref{eq:cWtilde}) have been referred to  as ``non-power expansions''. 

By construction, the series (\ref{eq:cW}) and  (\ref{eq:cWtilde}) must reproduce the low-order perturbative coefficients $F_{n}$ of the expansion (\ref{eq:F}), known from detailed calculations (Feynman diagrams or lattice QCD). If $F_n$ are known up to a certain order $N$, we can construct unambigously the expansions (\ref{eq:cW}) and  (\ref{eq:cWtilde}) truncated at the same order $N$. But, when reexpanded in powers of $a$,   these truncated expansions contain an infinite number of powers of $a$. The higher-order terms incorporate additional information about the expanded function, encoded in its singularities in the Borel plane. Thus, starting from a finite number of perturbative terms, we obtain a representation  that goes beyond standard finite-order perturbation theory.

A major difference between the standard perturbation expansion and the modified expansions concerns their analytic properties in the coupling complex plane. As we already mentioned, the exact QCD correlators  are expected to be singular at the origin of this plane. For the Adler function,  as proved in \cite{tHooft}, renormalization-group invariance and the multiparticle branch-points on the timelike axis of the momentum plane imply a complicated accumulation of singularities near the point $a=0$. Since the proof uses a
nonperturbative argument (the existence of multiparticle hadronic states),  it is not possible to see this feature in the standard perturbation theory:  the expansions in powers of the strong coupling $a$, truncated at finite orders, are holomorphic at the origin  of the  complex $a$ plane and cannot reproduce the singularity of the exact correlator at this point. 

For the non-power expansion functions  ${\cal W}_{n}(a)$ defined in (\ref{eq:Wn}), one expects a more complex structure in the $a$ plane, even after the regularization of the integral by the PV prescription. In \cite{Caprini:2001mn} it was shown, in the particular case of the Adler function,  that ${\cal W}_{n}(a)$    are analytic functions of real type, i.e. they satisfy the Schwarz reflection property  ${\cal W}_{n}(a^*)=  ({\cal W}_{n}(a))^*$, in the whole complex $a$ plane, except for a cut along the real negative axis $a<0$ and an essential singularity at $a=0$. Therefore, the expansion (\ref{eq:cW}), even if truncated at a finite order,  exhibits a feature of the full correlator, namely its singularity at the origin of the $a$ plane. The expansion functions  $\widetilde {\cal W}_{n}(a)$ defined in (\ref{eq:Wntilde}) have similar properties. 

 Since the expansion functions  have singularities at $a=0$,  their Taylor expansions around the origin will be  divergent series. By applying  Watson's
lemma \cite{Watson},\cite{Jeff}, it was shown in \cite{Caprini:2001mn} that ${\cal W}_{n}(a)$ can be expressed as
\beq\label{eq:Wna}
{\cal W}_{n}(a) = \sum\limits_{k=n}^N \xi_k^{(n)} k!\, a^k + M_n (N+1)!\, a^{N+1} + O\left({\rm e}^{-\frac{X}{ a}}\right), 
\eeq 
where $N$ is a positive integer, $M_n$ is independent of $N$, $X$ is an arbitrary positive parameter less than 1 and $\xi_k^{(n)}$ are defined by the Taylor expansions 
\begin{equation}\label{eq:wnser}
(\tilde w(u))^n=\sum\limits_{k=n}^\infty \xi_k^{(n)} u^k,\quad\quad n\ge 1.
\end{equation}
The representation (\ref{eq:Wna}) implies that the remainder
\beq\label{eq:Wnasim}
R_N^{(n)}\equiv{\cal W}_{n}(a) - \sum\limits_{k=n}^N \xi_k^{(n)} k!\, a^k
\eeq
satisfies the property
\beq\label{eq:Rn}
R_N^{(n)}=o(a^{N}),\quad \quad a\to 0_+, 
\eeq
which is precisely the definition of an asymptotic expansion \cite{Jeff}.

As shown in \cite{Caprini:2001mn}, the representation (\ref{eq:Wna}) is independent of the prescription adopted for the Borel-Laplace integral. The first term of each ${\cal W}_{n}(a)$ is
proportional to $n!\, a^n$ with a positive coefficient, preserving a
fundamental property of perturbation theory. But the series (\ref{eq:Wna}) 
are divergent: indeed, since the expansions (\ref{eq:wnser}) have the convergence
radii equal to 1, there are, for any $R > 1$, infinitely many $k$ such that  
$|\xi^{(n)}_k|>R^{-k}$ \cite{Jeff}. Actually, the fact that the series 
(\ref{eq:Wna}) is divergent is not surprising, in view of the singularities of the functions  ${\cal W}_{n}(a)$ at the origin of the $a$ plane. 
We conclude that, unlike the expansion functions $a^n$ of the standard perturbation theory, which are holomorphic at $a=0$,  the non-power expansion functions ${\cal W}_n(a)$ and $\widetilde {\cal W}_n(a)$ are singular at $a=0$ and admit divergent expansions in powers of $a$, being similar from this point of view to the exact correlators. 

On the other hand,  as proved in \cite{Caprini:2000js},\cite{Caprini:2001mn}, the new expansions  (\ref{eq:cW}) and  (\ref{eq:cWtilde}) have a tamed behaviour at high orders and, under certain conditions, they  may even converge. For instance, in the particular case of the Adler function in massless QCD,  if the coefficients $c_n$ of the expansion  (\ref{eq:cW}) satisfy the condition
\be\label{eq:cnb} |c_n| < C e^{\epsilon n^{1/2}}\ee
with $C>0$ for any $\epsilon >0$, the expansion (\ref{eq:cW}) converges in a domain of the $a$ complex plane. 
The condition depends on the specific positions $u_{UV}=-1$ and $u_{IR}=2$ of the first renormalons in the case of the Adler function. The proof presented in \cite{Caprini:2000js},\cite{Caprini:2001mn} can be easily generalized to other observables, by taking into account the positions of their leading renormalons. 

We emphasize that, although  a formal proof  is not available,  the validity of the condition (\ref{eq:cnb})  is expected to hold in full QCD. Moreover, the convergence has been investigated numerically and turned out to be very good  up to high orders in specific cases, as will be shown in Sect. \ref{sec:applic}. This property will play an important role in the discussion presented in the next section.

\section{Conformal mappings versus OPE}\label{sec:OPE}
Using the properties of the  expansions  based on the conformal mappings of the Borel plane, it was suggested in \cite{Caprini:2020lff} that they  may represent an  alternative to the standard OPE. In what follows, we briefly review the arguments supporting this conjecture.

In the frame of the standard OPE, proposed a long time ago in \cite{Shifman:1978bx}, the representation of the correlators contains, besides the truncated perturbative expansion,  a series of power corrections $\sim d_k/ \Lambda^{k}$, where $\Lambda$ is a relevant mass scale and the coefficients $d_k$ involve both perturbative factors, depending logarithmically on $\Lambda$, and nonperturbative condensates.  As we mentioned in the Introduction, in the language of hyperasymptotic theory, the power corrections are a first term of a transseries, i.e.  a sequence of truncated series, each of them  exponentially small in the expansion parameter of the previous one,  by which one can recover the expanded function from its divergent expansion. 

A first argument in favour of the conjecture formulated in \cite{Caprini:2020lff}  refers to  the analytic properties in the coupling plane. For OPE, the essential remark is that the powers $\Lambda^{-k}$ can be written approximately  as  $\exp [-k/(\beta_0 a(\Lambda^2))]$, where $a$ is the expansion parameter of the perturbative expansion,  calculated by solving the renormalization-group equation (\ref{eq:rge}). Therefore, the power corrections in the OPE can be identified with terms exponentially small in coupling, which are  nonanalytic at the origin of the complex $a$ plane.  On the other hand, as discussed in the previous section, the expansion functions ${\cal W}_n(a)$ and $\wt {\cal W}_n(a)$  exhibit intrinsic singularities near $a=0$.  Thus, both OPE and the  expansions based on the conformal mapping of the Borel plane incorporate a nonperturbative feature of the exact correlator, namely the singular behaviour at zero coupling (although neither of them can reproduce the complicated singularity structure near $a=0$ described in \cite{tHooft}). 

Another argument given in \cite{Caprini:2020lff} for the similarity of the two approaches concerns the behaviour in the complex Borel plane. The starting remark is that the divergent character of the perturbative expansion is due to the fact that the range of the Laplace-Borel integration (\ref{eq:Laplace}) extends  beyond the finite
radius of convergence of the expansion (\ref{eq:B}) of the Borel transform. In order to overcome this, the hyperasymptotic approach includes additional terms, which is equivalent in a certain sense to an analytic continuation of the Borel transform to the neighbourhood of the distant singularities \cite{Howls}. 
On the other hand, the expansions  (\ref{eq:Bw}) and  (\ref{eq:Bw1}), based on the conformal mapping of the Borel plane,  converge in the whole $u$ plane up to the cuts, achieving in a manifest way the analytic continuation outside the circle of convergence of the series (\ref{eq:B}). 
Therefore, the OPE and the method of conformal mapping represent complementary ways to effectively perform the analytic continuation in the Borel plane, in order to recover the expanded function.

A third argument is based on the fact that the nonperturbative terms in the OPE appear to depend on the form of the perturbative part. In particular, as discussed recently \cite{Ayala:2019uaw}, the  nonperturbative terms depend on the truncation order of the perturbation series. But, as discussed in the previous section,  the  expansions defined by the conformal mapping of the Borel plane,  even if truncated at finite orders, contain an infinite number of terms when reexpanded in powers of $a$. So, if they are supplemented by additional power corrections, the interpretation of these corrections in terms of condensates will be hard to give. 

Finally, an important argument is that the expansions based on conformal mappings of the Borel plane exhibit a tamed behaviour at large orders and have been shown to converge under some conditions (expected to be satisfied in QCD).  There are  no mathematical reasons for adding to a convergent series arbitrary terms having the same analyticity properties (one can add however terms with different analytic properties, if they are known to contribute, as we will discuss below).

Using the arguments reviewed above, it was conjectured in \cite{Caprini:2020lff} that the  expansions based on conformal mappings  provide by themselves a systematic way of recapturing the exact function,  being an alternative to the standard OPE.  This conjecture has  phenomenological consequences: in particular, as discussed in \cite{Caprini:2020lff}, when applied to the Adler function it leads to a more precise determination of the strong coupling $\alpha_s$ from the hadronic decays of the $\tau$ lepton. More detailed phenomenological tests of this conjecture are of course necessary,  using for instance suitable weighted integrals (moments) of the spectral function of the polarization amplitude up to a given energy $s_0$. These quantities are well described by standard perturbation theory and truncated OPE (see \cite{Boito:2019iwh} and references therein). Their alternative investigation  using the non-power expansions based on conformal mappings is of interest and will be considered in our future works.

We end this section with another remark of interest for phenomenological applications. As it is known, additional terms, which must supplement the power corrections of the  standard OPE, have been recently assumed to be present in the representation  of the physical correlators at large and moderate energies. These "duality-violating" (DV) contributions, discussed in several works \cite{Blok:1997hs},\cite{Shifman:2000jv},\cite{Cata:2005zj},\cite{Peris:2016jah},\cite{Boito:2017cnp},   decrease exponentially on the spacelike axis of the complex momentum plane and exhibit an oscillating behaviour when analytically continued to the timelike axis. One might ask whether such terms appear also in conjunction with the expansions based on the conformal mappings of the Borel plane. 

 As shown above,  both the power corrections in the OPE and the  expansions based on conformal mappings acount for singularities of the type  $\exp(-1/\alpha_s)$  at the origin of the coupling plane.  Mathematically, in the hyperasymptotic theory, the DV terms can be formally viewed as a further piece in a transseries, i.e. a series in powers of an expansion parameter exponentially small in the expansion parameter $\exp(-1/\alpha_s)$  of the OPE.  However, the concept of transseries cannot be applied directly to the expansions based on conformal mappings,  because they are not power series. One can nevertheless retain from the hyperasymptotic approach its implications on analyticity. The main idea is that in order to recover a function from its divergent perturbative series  one must add to the standard expansion in powers of the coupling (which are holomorphic at the expansion point) terms which are singular at the expansion point. Only this more general representation allows to describe nonperturbative features  of the unknown expanded function. In  QCD, as shown in \cite{Boito:2017cnp} using arguments independent of transseries, the hadronic resonances require singularities associated to both power corrections and exponentially-suppressed terms in the asymptotic expansions of the correlators.  Terms singular at the origin of the coupling plane, which can be associated to power corrections, are present in both the truncated OPE and the truncated non-power expansions based on the conformal mappings. It follows that the presence of additional terms singular at the origin of the coupling plane, which can be expressed as exponentially-suppressed terms  in the momentum plane, is natural in both approaches.  
 Therefore, DV contributions are expected to show up also in addition to the expansions based on the conformal mappings of the Borel plane, and should be taken into account in phenomenological analyses using this approach. 
\section{Applications}\label{sec:applic}
\subsection{Adler function in massless QCD}\label{sec:Adler}
We  consider the so-called reduced Adler function \cite{Beneke:2008ad}
\beq\label{eq:D}
\widehat{D}(s) \equiv 4 \pi^2 D(s) -1,
\eeq
where $D(s)=-s \,d\Pi(s)/ds$ is the logarithmic derivative of the invariant amplitude $\Pi(s)$ of the two-current correlation tensor and $s=q^2$ is the momentum variable.  From general principles of field theory, it is known that $\wh D(s)$ is an analytic function of real type (i.e. it satisfies the  Schwarz reflection property $\wh D(s^*)=\wh D^*(s)$) in the complex $s$ plane cut along the timelike axis for $s\ge 4 m_\pi^2$. 

 In QCD perturbation theory,  $\wh D(s)$ is written as an expansion  
\beq\label{eq:hatD}
\widehat{D}(s) =\sum\limits_{n\ge 1} [a(\mu^2)]^n \,
\sum\limits_{k=1}^{n} k\, c_{n,k}\, (\ln (-s/\mu^2))^{k-1},
\eeq
in powers of the renormalized strong coupling $a(\mu^2)$, defined in a certain renormalization scheme  at the renormalization scale $\mu$. 
The  coefficients  $c_{n,1}$ in (\ref{eq:hatD}) are obtained from the calculation of  Feynman diagrams, while  $c_{n,k}$ with $k>1$ are expressed in terms of  $c_{m,1}$ with $m< n$  and the perturbative coefficients $\beta_n$ of the $\beta$ function, defined in (\ref{eq:rge}).

For a large spacelike value $s<0$, one can choose in (\ref{eq:hatD}) the scale $\mu^2=-s$, and obtain the renormalization-group improved expansion
\beq\label{eq:hatD1}
\widehat{D}(s) =\sum\limits_{n\ge 1} c_{n,1}\, [a(-s)]^n,
\eeq
where $a(-s)\equiv \alpha_s(-s)/\pi$ is the running coupling. The expansions (\ref{eq:hatD})  and (\ref{eq:hatD1}) are often used also for complex values of $s$, outside the timelike axis $s>0$.

The Adler function has been calculated in the $\overline{{\rm MS}}$ scheme to order $\alpha_s^4$  (see \cite{Baikov:2008jh} and references therein). For $n_f=3$, the leading coefficients $c_{n,1}$  have the values:
\be\label{eq:cn1}
c_{1,1}=1,\,\, c_{2,1}=1.640,\,\, c_{3,1}=6.371,\,\, c_{4,1}=49.076.
\ee
Estimates of the next coefficient $c_{5,1}$ are also available, the most recent determinations being reported in  \cite{Caprini:2019kwp},\cite{Boito:2018rwt},\cite{Jamin:2021qxb}.

At high orders $n$, the coefficients increase factorially, more exactly $c_{n,1}\approx K\,b^n n !\, n^{c}$,  where  $K$, $b$ and $c$ are  constants  \cite{Beneke:1998ui}. This behaviour is encoded in the singularities of the Borel transform $B_{\widehat{D}}(u)$, defined as in (\ref{eq:B}): this function  has  cuts along the real axis in the $u$ plane, the position of the first IR and UV renormalons being at
\beq\label{eq:UVIR}
u_{UV}=-1, \quad\quad u_{IR}=2.
\eeq 
 The nature of the first branch points of  the Borel transform  has been discussed in \cite{Beneke:2008ad},  where the following expressions are given
\beq\label{eq:gamma}
\gamma_{UV}=2-\frac{\beta_1}{\beta_0^2},\quad\quad \gamma_{IR}=1+2\,\frac{\beta_1}{\beta_0^2},
\eeq
in terms of the first coefficients of the $\beta$ function,  defined in (\ref{eq:beta01}). We note that while $\gamma_{IR}$ is determined from renormalization-group invariance, the expression of $\gamma_{UV}$ is a reasonable assumption,
based on the knowledge of some of the anomalous dimensions involved
and on the results in the large-$\beta_0$ limit.

In our analysis, we shall use (\ref{eq:gamma}) 
for $n_f=3$,  when 
\begin{equation}\label{eq:gamma12}
\gamma_{UV} = 1.21,    \quad\quad   \gamma_{IR} = 2.58. 
\end{equation}

The perturbative expansions of the Adler function based on the optimal conformal mapping (\ref{eq:w}) of the Borel plane are obtained from the general expressions (\ref{eq:cW})-(\ref{eq:Sw}) given in Sect. \ref{sec:conf}, using the particular values (\ref{eq:UVIR}) and (\ref{eq:gamma12}). These expansions have been proposed for the first time in \cite{Caprini:1998wg}.  Further investigations and applications are reported in  \cite{Caprini:2000js}-\cite{Caprini:2020lff}. In particular, the new expansions have been used for increasing the accuracy of the extraction of the strong coupling $\alpha_s$ from hadronic $\tau$ decays (for a recent determination, see  \cite{Caprini:2020lff}). 

In the present paper we focus on the convergence properties of the expansions (\ref{eq:cW}) and (\ref{eq:cWtilde}), which are an important argument for the conjecture discussed in the previous section. The convergence of the new expansions has been investigated on models of the Adler function  inspired from QCD in \cite{Caprini:2009vf},\cite{Caprini:2011ya},\cite{Caprini:2019kwp}.
These models, proposed for the first time in \cite{Beneke:2008ad} and considered in many other studies (for a recent discussion see \cite{Jamin:2021qxb}), parametrize the Borel transform $B_{\wh D}(u)$  as a finite sum of IR and UV renormalons and a regular part,  satisfy renormalization-group invariance and reproduce the known low-order coefficients (\ref{eq:cn1}) of the expansion (\ref{eq:hatD1}). The exact value of the Adler function is obtained by inserting the Borel transform described by each model in the PV-regulated Borel-Laplace integral (\ref{eq:Laplace}). 

\begin{table}[thb]
\caption{Adler function of the ``reference model'' \cite{Beneke:2008ad}, calculated with the standard perturbative  expansion (\ref{eq:Dtrunc}) and the non-power expansions (\ref{eq:cW})-(\ref{eq:Sw}), for various truncation orders $N$. For each column we indicate the specific equations used in the calculations. Exact value: $\wh D(-m_\tau^2)=0.137706$.} \vspace{0.15cm}
\label{tab:1}
 \renewcommand{\tabcolsep}{0.55pc} 
\renewcommand{\arraystretch}{1.15} 
\begin{tabular}{|l|c|c| c | c| }\hline\hline
$N$ \,\,\,    & Eq. (\ref{eq:Dtrunc})  &\,\,Eqs. (\ref{eq:cW}), (\ref{eq:Wn})& Eqs. (\ref{eq:cWtilde}), (\ref{eq:Wntilde}),  (\ref{eq:Su})&Eqs. (\ref{eq:cWtilde}), (\ref{eq:Wntilde}),  (\ref{eq:Sw}) \\\hline
10 & 0.155429 & 0.142247& 0.137679& 0.137763       \\   
 11 & 0.149068  &0.139757& 0.137703& 0.137733     \\
 12 & 0.191213 & 0.137235&0.137652& 0.137700   \\
 13 & 0.114491 & 0.135647& 0.137648&0.137712     \\
 14 & 0.417809& 0.135401& 0.137662 &0.137729    \\
 15 & -0.442007& 0.136258& 0.137637 & 0.137724   \\
 16 & 2.80676& 0.137549& 0.13765 &0.137715     \\
 17 & -8.76330 & 0.138553& 0.137681 & 0.137714   \\
 18 & 37.9988 & 0.138851& 0.137682& 0.137716     \\
 19 & -154.7999& 0.138470&0.137689 & 0.137714     \\
 20 & 700.409 & 0.137788& 0.137703 & 0.137711   \\
 21 & -3248.105 & 0.137259& 0.137705 & 0.137709     \\
 22 & 15993.08 & 0.137139& 0.137707 & 0.137709    \\
 23 & -81886.8 & 0.137384 &0.137711 &0.137709     \\
 24 & 439277.8 & 0.137744&0.137710& 0.137707     \\
 25  &-2.45 $\times 10^6$&0.137973 &0.137708& 0.137706     \\[0.04cm]
\hline\hline \end{tabular}\end{table}

We consider for illustration the perturbative calculation of the Adler function on the spacelike axis, using  the so-called ``reference model'' proposed in  \cite{Beneke:2008ad} (see \cite{Caprini:2019kwp} for details on this and other models).  From the perturbative
coefficients of these models, calculated exactly to any order, one can obtain the standard perturbation expansion  (\ref{eq:hatD1}) and construct also the improved ones, given in Eqs. (\ref{eq:cW})-(\ref{eq:Sw}). 

As an example, in Table \ref{tab:1} we present the results for the Adler function at the spacelike point $s=-m_\tau^2$, calculated for $\alpha_s(m_\tau^2)=0.32$.  We give  the values of the truncated standard expansion
\beq\label{eq:Dtrunc}
\wh D(-m_\tau^2)=\sum\limits_{n=1}^N c_{n,1}\, [a(m_\tau^2)]^n,
\eeq
and the improved expansions  (\ref{eq:cW}) and (\ref{eq:cWtilde}) truncated at the same  order. The exact value calculated in the model is  $\wh D(-m_\tau^2)=0.137706$.

Since we are interested in comparing the large-order behaviour, only the results for truncation orders $N\ge 10$ are shown. One can see the divergent pattern of the standard expansion given in column 2, and the good convergence of the expansions (\ref{eq:cW}) and (\ref{eq:cWtilde}). As seen from the last two columns, the inclusion of the softening factor $S(u)$, of the form (\ref{eq:Su}) or  (\ref{eq:Sw}),  improves the precision at both moderate and high orders.  This pattern is preserved to higher orders: for 
instance, for $N=40$, the standard expansion gives $2.39 \times 10^{19}$, while the improved expansion (\ref{eq:cWtilde}) with the softening factor (\ref{eq:Sw})  reproduces the exact value to 7 digits.  

\subsection{Self-energy of static sources}\label{sec:static}
As a second example, we consider the perturbative expansions of the energy of static sources, calculated  on the lattice. Perturbative coefficients for sources in both the fundamental and adjoint representations in SU(3) gluodynamics have been obtained  in \cite{Bali:2013pla},\cite{Bali:2013qla}. Here we will consider the self-energy of a static quark, which is relevant for the pole mass calculation on the lattice. It is 
expanded as
\beq\label{eq:E}
E(\alpha)=\sum_{n=0}^{\infty} c_n \alpha^{n+1},
\eeq
where the coefficients $c_n$ are given to order $\alpha^{20}$  in \cite{Bali:2013pla},\cite{Bali:2013qla}.

The Borel transform $B_E(u)$ is defined as in (\ref{eq:B}), using the coefficients $b_n$ 
\beq
b_n=\frac{ \pi^{n+1} c_n}{\beta_0^n n!},
\eeq
obtained from (\ref{eq:bn}), with $F_n$ replaced by $c_n \pi^{n+1}$.

The asymptotic growth of the coefficients $c_n$ indicate the presence of a dominant IR renormalon at $u=1/2$ in the Borel transform $B_E(u)$. In \cite{Ayala:2019hkn}, arguments in favour of a UV renormalon at $u=-1$ for the pole mass have been given. As it is known, the quark self-energy and the pole mass have the same leading IR renormalon, but the subleading renormalons can be different. So, the presence or the absence of a subleading UV renormalon at $u=-1$ for the energy of a static quark is an open question. 

 The improved expansions of $E(\alpha)$  are obtained, as discussed in Sect. \ref{sec:conf}, by expanding the Borel transform $B_E(u)$ in powers of  the optimal conformal variable $\tilde w(u)$ defined in (\ref{eq:w}), which requires the positions of the first IR and UV renormalons. 
In the present case, we used the knowledge on the first IR renormalon and two assumptions on the UV renormalon, taking  in (\ref{eq:w}) either
\beq\label{eq:stIR}
u_{IR}=1/2, \quad\quad u_{UV}\to \infty,
\eeq
which corresponds to the absence of UV renormalons, or
\beq\label{eq:stIRUV}
u_{IR}=1/2, \quad\quad u_{UV}=-1,
\eeq
which corresponds to the presence of a UV renormalon $u=-1$.

As discussed in Sect. \ref{sec:conf}, the precision at moderate orders of the expansion based on conformal mappings  is improved by implementing the known behaviour of the Borel transform near the leading singularities. As shown in  \cite{Bali:2013pla},\cite{Bali:2013qla}, in the present case only the behaviour near the branch point at $u_{IR}$ is known. Therefore, we took the factor $S(u)$ appearing in the expansion functions (\ref{eq:Wntilde})  of the form
\be\label{eq:Sust}
S(u)=(1-u/u_{IR})^{\gamma_{IR}}, 
\ee
or
\be\label{eq:Swst}
S(u)=(1-\tilde w(u))^{2\gamma_{IR}}.
\ee
Here the exponent $\gamma_{IR}$ is expressed as \cite{Bali:2013pla}
\beq\label{eq:gammaIR}
\gamma_{IR}=1+\frac{\beta_1}{2 \beta_0^2}
\eeq
in terms of the coefficients given in (\ref{eq:beta01}). Taking $n_f=0$, as in  \cite{Bali:2013pla}, we obtain the numerical value  $\gamma_{IR}=1.42$. We note that in this case the residue $r_{IR}$ is also known \cite{Bali:2013pla},\cite{Bali:2013qla}. This information can be easily implemented in the formalism. However, for simplicity, we did not include it in the exploratory study reported here.

In Table \ref{tab:2} we give the values of the truncated standard expansion 
\beq\label{eq:Etrunc}
E(\alpha)=\sum_{n=0}^{N} c_n \alpha^{n+1},
\eeq
and of the  expansions obtained from (\ref{eq:cWtilde}), truncated at the same order. For each choice, (\ref{eq:stIR}) or  (\ref{eq:stIRUV}), of the parameters specifying the conformal mapping (\ref{eq:w}), we consider both expressions (\ref{eq:Sust}) and (\ref{eq:Swst}) of the softening factor.  In the calculations, we used as input the central values of the perturbative coefficients given in Table 1 of   \cite{Bali:2013qla},  and the standard value of $\alpha$ adopted in lattice calculations, given by $\beta\equiv 6/g^2=3/(2 \pi \alpha) \approx 6$. 

\begin{table}[thb]
\caption{Static quark energy  calculated with the standard  perturbative expansion  (\ref{eq:Etrunc})  and the non-power expansions (\ref{eq:cWtilde}) for various truncation orders $N$. For each column we indicate the specific equations used in the calculations.} \vspace{0.15cm}
\label{tab:2}
 \renewcommand{\tabcolsep}{0.55pc} 
\renewcommand{\arraystretch}{1.15} 
\begin{tabular}{|l|c|c| c | c|c| }\hline\hline
$N$ & Eq. (\ref{eq:Etrunc}) & Eqs. (\ref{eq:stIR}), (\ref{eq:Sust})& Eqs. (\ref{eq:stIR}),  (\ref{eq:Swst})& Eqs. (\ref{eq:stIRUV}), (\ref{eq:Sust}) & Eqs. (\ref{eq:stIRUV}), (\ref{eq:Swst})  \\\hline
 2 & 0.284735 & 0.350487& 0.344406& 
  0.347062 & 0.341147 \\
 3& 0.317278 & 0.347779 & 0.358333 & 
  0.353279 & 0.359863 \\
 4& 0.344174 & 0.356579 & 0.354036 & 
  0.351390 & 0.356929 \\
 5& 0.368551 & 0.365017 & 0.365037& 
  0.361116 & 0.359365 \\
 6& 0.392622 & 0.365181 & 0.367351 & 
  0.366431 & 0.368159 \\
 7& 0.418543 & 0.372964 & 0.370863 & 
  0.366210 & 0.368827 \\
 8& 0.449091 & 0.373185 & 0.374439 & 
  0.371689 & 0.370646 \\
 9& 0.488648 & 0.375564 & 0.374692 & 
  0.373280 & 0.375383 \\
 10& 0.544997 & 0.373381 & 0.380496 & 
  0.374640 & 0.375210 \\
 11& 0.633027 & 0.384267 & 0.382299 & 
  0.379260 & 0.379305 \\
 12& 0.782726 & 0.388539 & 0.383148 & 
  0.378938 & 0.382176 \\
 13& 1.059363 & 0.269775 & 0.270466 & 
  0.397753 & 0.392178 \\
 14& 1.610702 & -0.538473 & 0.633450 & 
  0.387688 & 0.374793 \\
 15& 2.790082 & 1.785543 & 3.967640 & 
  0.381146 & 0.374410 \\
 16& 5.478731 & 16.84796 & 3.711328 & 
  0.368613 & 0.363885 \\
 17& 11.99994 & 16.98481 & -24.38370 & 
  0.379323 & 0.350790 \\
 18& 28.71730 & -99.12591 & -60.85833 & 
  0.214992 & 0.202010 \\
 19& 73.91183 & -291.2176 & 92.54484 & 
  0.172661 & 0.393075 
\\[0.04cm]
\hline\hline \end{tabular}\end{table}

The values given in column 2 illustrate the divergent pattern of the standard expansion at large orders. The results presented in columns 3 and 4 exhibit a better stability at moderate orders. However, at large orders the expansions start to show a divergent behaviour. This behaviour can be explained by the presence of a UV renormalon at $u=-1$, which was not taken into account by the conformal mapping. Indeed, with the choice (\ref{eq:stIR}), the conformal mapping (\ref{eq:w}) takes the particular form (\ref{eq:wIR}), and  maps the $u$ plane cut only along $u\ge 1/2$ on the unit disk in the $w$ plane. By this mapping, the point $u=-1$ becomes $w=-0.268$. Therefore, if the Borel transform $B_E(u)$ has a singularity at $u=-1$, the expansion of this function in powers of the variable $w$ will converge only in the disk $|w|<0.268$, while the Borel-Laplace integration (\ref{eq:Laplace}) extends outside this region. This can explain the divergent pattern of the expansions (\ref{eq:cWtilde}) in this case.

 By contrast, with the choice (\ref{eq:stIRUV}), the function $\tilde w(u)$ defined in (\ref{eq:w}) maps
 the $u$ plane cut along $u\le -1$ and $u\ge 1/2$ onto the unit disk $|w|<1$. Therefore, even if the Borel transform $B_E(u)$ has a singularity at $u=-1$,  its expansion in powers of $w$ will converge in the whole $u$ plane, up to the cuts. The good convergence of the expansions shown in the last two columns of Table \ref{tab:2} indicate that (\ref{eq:stIRUV}) is the good choice of the conformal mapping in this case.

The conclusion is that the expansions based on the conformal mapping of the Borel plane provide arguments in favour of a UV renormalon at $u=-1$ in the Borel transform $B_E(u)$  of the self-energy of a static quark. 
Of course, this is a preliminary result which needs further investigations and confirmation. 

\section{Summary and conclusions}\label{sec:conc}
In the present work we reviewed the application of the method of series acceleration by conformal mappings  to perturbative QCD. Mathematically, the method amounts to reordering a power series as an expansion in powers of a new variable, which performs the conformal mapping of a part of the holomorphy domain of the expanded function, containing the expansion point,  onto the unit disk in a new complex plane. This ensures a larger domain of convergence and an increased convergence rate of the new expansion, compared to the original one.

In order to apply the method, the expanded function must be analytic in a region around the expansion point. This condition is not satisfied by the standard perturbative expansions of the QCD correlators in powers of the coupling $a$, which are in many cases singular at $a=0$.  However, the method can be applied to the Borel transforms of the correlators, since their singularities in the Borel plane, which encode the large-order increase of the perturbative coefficients, are placed at a finite distance from the origin $u=0$. 

As shown a long time ago \cite{CiFi}, one can find an optimal variable,  such that the  expansion in powers of that variable  converges in the whole complex plane and has the best asymptotic convergence rate. As shown in \cite{CiFi}, this variable achieves the  conformal mapping of the whole analyticity domain of the expanded function onto a unit disk. We refer to this as to the ``optimal conformal mapping'' for series acceleration. 

In QCD, the optimal conformal mapping for improving the perturbation series  was written down for the first time in \cite{Caprini:1998wg}, for the particular case of the Adler function. It depends on the specific positions of the leading IR and UV renormalons of the corresponding Borel transform. In the present paper, we gave in Sect. \ref{sec:conf} the most general expression of the optimal mapping $\tilde w(u)$, for the generic case of a QCD observable with cuts along the real axis of the Borel plane due to IR and UV renormalons. In the same section, we defined  in Eqs. (\ref{eq:cW})-(\ref{eq:Sw}) the new, non-power expansions of the observable, taking into account the information about the leading  singularities in the Borel plane.

In Sect. \ref{sec:prop} we summarized the main properties of the new expansions, emphasizing the fact that the expansion functions have nonperturbative features much like the expanded function itself. In particular, the expansion functions  ${\cal W}_n(a)$ and $\widetilde {\cal W}_n(a)$  exhibit a singularity at $a=0$ and their expansions in powers of $a$ are divergent. On the other hand, the new expansions  (\ref{eq:cW}) and  (\ref{eq:cWtilde}) have a more tamed behaviour at large orders, and may converge under certain conditions.
 
In  Sect. \ref{sec:OPE}  we reviewed the arguments presented in \cite{Caprini:2020lff}, where it was suggested that the expansions based on the conformal mapping of the Borel plane might be an alternative to the OPE for recapturing nonperturbative features of the exact QCD correlators. We note that the possibility of the resurgence of the Adler function without adding power corrections has been discussed recently also in \cite{Maiezza:2021mry}. 

Finally, in Sect. \ref{sec:applic} we briefly discussed the application of the method for two specific observables: the Adler function in massless QCD and the energy of a static quark. In both cases, we focused on illustrating the tamed behaviour of the new expansions  at high orders, which is an important argument in favour of the conjecture discussed in Sect. \ref{sec:OPE}. In the case of the Adler function, for which many applications of the conformal mappings  already exist \cite{Caprini:1998wg}-\cite{Caprini:2020lff},
we  illustrated the good convergence of the new expansions for the Adler function calculated on the spacelike axis in the frame of a realistic model inspired from QCD. For the energy of the static quark, we used as input the perturbative coefficients calculated on the lattice in \cite{Bali:2013pla},\cite{Bali:2013qla}, and examined the large-order behaviour of the expansions based on two conformal mapings,  assuming that a nondominant UV renormalon is either absent or present. We emphasize that the present analysis is the first application of the method of conformal mappings to this observable. The preliminary results show that the perturbative coefficients of the quark self-energy seem to feel the presence of a singularity at $u=-1$, besides the leading singularity at $u=1/2$ in the Borel plane. The application of the conformal mappings to other observables calculated on the lattice and the comparison with the hyperasymptotic expansions \cite{Ayala:2019uaw} are objectives of our future work.

\vspace{0.3cm}
\noindent
{\bf Acknowledgments}~ I am grateful to A. Pineda for very useful discussions. This work was supported by the Romanian Ministry of Research, Innovation and Digitization,  Contract PN 19060101.



\begin{thebibliography}{}
\bibitem{Dyson:1952tj}
F.~J. Dyson, Divergence of perturbation theory in quantum
  electrodynamics, \href{https://doi.org/10.1103/PhysRev.85.631}{{Phys.
  Rev. {\bfseries 85}, 631 (1952)}}


\bibitem{tHooft} G. 't Hooft, Can we make sense out of Quantum Chromodynamics?
in \href{https://link.springer.com/book/10.1007/978-1-4684-0991-8}{{\it The Whys of Subnuclear Physics}}, edited by A. Zichichi (Plenum Press, New York, 1979), p. 943-982

\bibitem{Lautrup:1977hs} 
  B.~E.~Lautrup, On high order estimates in QED, \href{https://www.sciencedirect.com/science/article/abs/pii/0370269377901459?via%3Dihub}{Phys.\ Lett.\  {\bf 69B}, 109 (1977)}
 
\bibitem{Broadhurst:1992si}
D.~J. Broadhurst, Large N expansion of QED: Asymptotic photon propagator
  and contributions to the muon anomaly, for any number of loops,
  \href {https://doi.org/10.1007/BF01560355} {Z. Phys. {\bf C58}, 339 (1993)}

\bibitem{Beneke:1994qe}
M.~Beneke and V.~M. Braun, Naive nonabelianization and resummation of fermion bubble chains,
  \href{https://doi.org/10.1016/0370-2693(95)00184-M}{Phys. Lett. {\bf B348}, 513 (1995)},
  \href{https://arxiv.org/abs/hep-ph/9411229}{arXiv:hep-ph/9411229}

\bibitem{Beneke:1992ch}
M.~Beneke, Large order perturbation theory for a physical quantity, \href{https://doi.org/10.1016/0550-3213(93)90554-3}{Nucl. Phys. {\bf B405}, 424 (1993)}

\bibitem{Beneke:1998ui}
M.~Beneke, Renormalons,
  \href{https://doi.org/10.1016/S0370-1573(98)00130-6}{{Phys. Rept.}
  {\bf 317}, 1 (1999)},  \href{https://arxiv.org/abs/hep-ph/9807443}{arXiv:hep-ph/9807443}

\bibitem{Bauer:2011ws} C. Bauer, G. S. Bali and A. Pineda,  Compelling evidence of renormalons in QCD from high
order perturbative expansions, \href{https://journals.aps.org/prl/abstract/10.1103/PhysRevLett.108.242002}{Phys. Rev. Lett. {\bf 108},  242002 (2012)}, \href{https://arxiv.org/abs/1111.3946}{arXiv:1111.3946}

\bibitem{Mueller1985} A. H. Mueller, On the structure of infrared renormalons in physical processes at high energies, \href{https://www.sciencedirect.com/science/article/pii/0550321385904857?via%3Dihub}{Nucl. Phys. {\bf B250}, 327 (1985)}

\bibitem{Mueller:1993pa} 
  A.~H.~Mueller, Combining higher twist terms with finite order perturbative contributions, \href{https://www.sciencedirect.com/science/article/abs/pii/037026939391297Z?via%3Dihub}{{Phys.\ Lett.\ B {\bf 308}, 355 (1993)}}

\bibitem{Shifman:1978bx} 
  M.~A.~Shifman, A.~I.~Vainshtein and V.~I.~Zakharov, QCD and resonance physics, 
\href{https://www.sciencedirect.com/science/article/pii/0550321379900221?via%3Dihub}{Nucl.\ Phys.\ B {\bf 147}, 385 (1979)}, \href{https://www.sciencedirect.com/science/article/pii/0550321379900233?via%3Dihub}{{\bf 147}, 448 (1979)}

\bibitem{BerryHowls} M.V. Berry and C.J. Howls, Hyperasymptotics, \href{https://royalsocietypublishing.org/doi/abs/10.1098/rspa.1990.0111}{Proceedings of the Royal  Society A: Mathematical, Physical and Engineering Sciences {\bf 439}, 653 (1990)}

\bibitem{Howls} C.J. Howls, An introduction to hyperasymptotics using  Borel-Laplace transforms, in the series \href{https://repository.kulib.kyoto-u.ac.jp/dspace/handle/2433/60642}{Algebraic Analysis of Singular Perturbations}, Kyoto Univ. (1996)

\bibitem{Dorigoni:2014hea}  D.~Dorigoni, An introduction to resurgence, transseries and alien calculus, \href{https://www.sciencedirect.com/science/article/abs/pii/S0003491619301691?via%3Dihub}{Annals of Phys.\  {\bf 409}, 167914 (2019)}, \href{https://arxiv.org/abs/1411.3585}{{arXiv:1411.3585}}

\bibitem{CiFi} S. Ciulli and J. Fischer, A convergent set of integral equations for singlet proton-proton scattering, \href{https://www.sciencedirect.com/science/article/abs/pii/0029558261904138?via%3Dihub} {Nucl. Phys. {\bf 24}, 465 (1961)}

\bibitem{Frazer} W. R. Frazer, Applications of conformal mapping to the phenomenological representation of scattering amplitudes, \href{https://journals.aps.org/pr/abstract/10.1103/PhysRev.123.2180}{Phys. Rev. {\bf 123}, 2180 (1961)}

\bibitem{Seznec:1979ev}  R. Seznec and J. Zinn-Justin, 	
Summation of divergent series by order dependent mappings: application to the anharmonic oscillator and critical exponents in field theory, \href{https://aip.scitation.org/doi/10.1063/1.524247}{J. Math. Phys. {\bf 20}, 1398 (1979)}

\bibitem{ZinnJustin:2010ng}    J. Zinn-Justin and U. D. Jentschura,  
Order-dependent mappings: strong coupling behaviour from weak coupling expansions in non-Hermitian theories, \href{https://aip.scitation.org/doi/10.1063/1.3451104} {J. Math. Phys. {\bf 51}, 072106 (2010)}, \href{https://arxiv.org/abs/1006.4748}{arXiv:1006.4748}

\bibitem{Altarelli:1994vz} G.~Altarelli, P.~Nason and G.~Ridolfi, A study of ultraviolet renormalon ambiguities in the determination of $\alpha_s$ from $\tau$ decay, \href{https://link.springer.com/article/10.1007%2FBF01566673}{Z.\ Phys.\ C {\bf 68}, 257 (1995)}, \href{https://arxiv.org/abs/hep-ph/9501240}{arXiv:hep-ph/9501240}

\bibitem{Caprini:1998wg}
  I. Caprini and J. Fischer, Accelerated convergence of perturbative QCD by optimal conformal mapping of the Borel plane, \href{https://journals.aps.org/prd/abstract/10.1103/PhysRevD.60.054014} {{Phys.\ Rev.}\  D {\bf 60}, 054014 (1999)}, \href{https://arxiv.org/abs/hep-ph/9811367}{arXiv:hep-ph/9811367}


\bibitem{Caprini:2000js} 
  I. Caprini and J. Fischer, Convergence of the expansion of the Laplace-Borel integral in perturbative QCD improved by conformal mapping, \href{https://journals.aps.org/prd/abstract/10.1103/PhysRevD.62.054007}{ Phys.\ Rev.\  D {\bf 62}, 054007 (2000)}, \href{https://arxiv.org/abs/hep-ph/0002016}{arXiv: hep-ph/0002016}

\bibitem{Caprini:2001mn} I. Caprini and J. Fischer,  Analytic continuation and perturbative expansions in QCD, 
\href{https://link.springer.com/article/10.1007%2Fs100520100880}{Eur.\ Phys.\ J.\  C {\bf 24}, 127 (2002)}, \href{https://arxiv.org/abs/hep-ph/0110344}{arXiv:hep-ph/0110344}
  
\bibitem{Cvetic:2001sn} 
  G.~Cvetic and T.~Lee, Bilocal expansion of Borel amplitude and hadronic $\tau$ decay width, \href{https://journals.aps.org/prd/abstract/10.1103/PhysRevD.64.014030}
  {Phys.\ Rev.\ D {\bf 64}, 014030 (2001)}, \href{https://arxiv.org/abs/hep-ph/0101297}{arXiv:hep-ph/0101297}

\bibitem{Jeong:2002ph} 
  K.~S.~Jeong and T.~Lee, Estimating higher order perturbative coefficients using Borel transform, \href{https://www.sciencedirect.com/science/article/pii/S0370269302029763?via%3Dihub}{Phys.\ Lett.\ B {\bf 550}, 166 (2002)},
 \href{https://arxiv.org/abs/hep-ph/0204150}{arXiv:hep-ph/0204150}

\bibitem{Caprini:2009vf} I. Caprini and J. Fischer,  $\alpha_s$ from $\tau$ decays: Contour-improved versus fixed-order summation in a new QCD perturbation expansion, \href{https://link.springer.com/article/10.1140%2Fepjc%2Fs10052-009-1142-8}{{Eur. Phys. J. C {\bf 64}, 35  (2009)}}, 
\href{https://arxiv.org/abs/0906.5211}{arXiv:0906.5211}


\bibitem{Caprini:2011ya}
I.~Caprini and J.~Fischer, Expansion functions in perturbative QCD and the determination of $\alpha_s(M_\tau^2)$,
  \href{https://doi.org/10.1103/PhysRevD.84.054019}{Phys. Rev. D {\bf 84},  054019 (2011)}, \href{https://arxiv.org/abs/1106.5336}{arXiv:1106.5336}

\bibitem{Abbas:2012fi} 
  G.~Abbas, B.~Ananthanarayan, I.~Caprini, and J.~Fischer, Perturbative expansion of the QCD Adler function improved by renormalization-group summation and analytic continuation in the Borel plane,  \href{https://journals.aps.org/prd/abstract/10.1103/PhysRevD.87.014008}{Phys.\ Rev.\ D {\bf 87},  014008   (2013)}, \href{https://arxiv.org/abs/1211.4316}{arXiv:1211.4316}

\bibitem{Abbas:2013usa}
G.~Abbas, B.~Ananthanarayan, I.~Caprini and J.~Fischer, Expansions of
  $\tau$ hadronic spectral function moments in a nonpower QCD perturbation theory with tamed large order behaviour, \href{https://doi.org/10.1103/PhysRevD.88.034026}{Phys. Rev. D {\bf 88},  034026  (2013)}, \href{https://arxiv.org/abs/1307.6323}{arXiv:1307.6323}


\bibitem{Caprini:2018agy}
I.~Caprini, Renormalization-scheme variation of a QCD perturbation
  expansion with tamed large-order behaviour, \href{https://doi.org/10.1103/PhysRevD.98.056016}{Phys. Rev. D
  {\bf 98}, 056016  (2018)}, \href{https://arxiv.org/abs/1806.10325}{arXiv:1806.10325}

\bibitem{Caprini:2019kwp} 
  I.~Caprini, Higher-order perturbative coefficients in QCD from series acceleration by conformal mappings,
 \href{https://journals.aps.org/prd/abstract/10.1103/PhysRevD.100.056019}{Phys.\ Rev.\ D {\bf 100}, 056019 (2019)}, \href{https://arxiv.org/abs/1908.06632}{arXiv:1908.06632}

\bibitem{Caprini:2020lff}
I.~Caprini, Conformal mapping of the Borel plane: going beyond perturbative QCD,
\href{https://journals.aps.org/prd/abstract/10.1103/PhysRevD.102.054017}{Phys. Rev. D {\bf 102},  054017 (2020)},
\href{https://arxiv.org/abs/2006.16605}{arXiv:2006.16605}

\bibitem{Caprini:1999ma} 
  I.~Caprini and M.~Neubert, Borel summation and momentum plane analyticity in perturbative QCD, 
\href{https://iopscience.iop.org/article/10.1088/1126-6708/1999/03/007}{JHEP {\bf 03}, 007 (1999)}, 
 \href{https://arxiv.org/abs/hep-ph/9902244}{arXiv:hep-ph/9902244}

\bibitem{Watson}
G.N. Watson, A theory of asymptotic series, \href{https://royalsocietypublishing.org/doi/abs/10.1098/rsta.1912.0007}{{Philos. Trans. Roy. Soc. London, Series A  {\bf 211},  279--313 (1912)}}

\bibitem{Jeff} H. Jeffreys, {\it Asymptotic Approximations}, Clarendon Press, Oxford (1962)

\bibitem{Ayala:2019uaw}
C.~Ayala, X.~Lobregat and A.~Pineda, Superasymptotic and hyperasymptotic approximation to the operator product expansion, \href{https://journals.aps.org/prd/abstract/10.1103/PhysRevD.99.074019}{Phys. Rev. D \textbf{99},  074019 (2019)}, \href{https://arxiv.org/abs/1902.07736}{arXiv:1902.07736}

\bibitem{Boito:2019iwh}
D.~Boito, M.~Golterman, K.~Maltman and S.~Peris,
Evidence against naive truncations of the OPE from $e^+e^- \to$ hadrons below charm,
\href{https://journals.aps.org/prd/abstract/10.1103/PhysRevD.100.074009}{Phys. Rev. D \textbf{100}, 074009 (2019)}, \href{https://arxiv.org/abs/1907.03360}{arXiv:1907.03360}

\bibitem{Blok:1997hs} 
  B.~Blok, M.~A.~Shifman, and D.~X.~Zhang, An illustrative example of how quark hadron duality might work,
  \href{https://journals.aps.org/prd/abstract/10.1103/PhysRevD.57.2691}{{Phys.\ Rev.\ D {\bf 57}, 2691 (1998)}}
 \href{https://journals.aps.org/prd/abstract/10.1103/PhysRevD.59.019901}{{{\bf 59}E, 019901 (1999)}},
  \href{https://arxiv.org/abs/hep-ph/9709333} {arXiv:hep-ph/9709333}.

\bibitem{Shifman:2000jv}
M.~A. Shifman, Quark-hadron duality,  in  \href{https://doi.org/10.1142/9789812810458_0032}{\it At the frontier of particle physics},  pp.~1447--1494, (World Scientific, Singapore, 2001),
  \href{https://arxiv.org/abs/hep-ph/0009131}{arXiv:hep-ph/0009131}

\bibitem{Cata:2005zj}
O.~Cat{\`a}, M.~Golterman and S.~Peris, {{Duality violations and spectral
  sum rules}}, \href{https://doi.org/10.1088/1126-6708/2005/08/076}{{JHEP}
  {\bf 08}, 076  (2005)},
  \href{https://arxiv.org/abs/hep-ph/0506004}{arXiv:hep-ph/0506004}

\bibitem{Peris:2016jah}
S.~Peris, D.~Boito, M.~Golterman and K.~Maltman, {{The case for duality
  violations in the analysis of hadronic $\tau$ decays}},
  \href{https://doi.org/10.1142/S0217732316300317}{Mod. Phys. Lett A.
  {\bf 31}, 1630031  (2016)}, \href{https://arxiv.org/abs/1606.08898}{arXiv:1606.08898}.

\bibitem{Boito:2017cnp}
D.~Boito, I.~Caprini, M.~Golterman, K.~Maltman and S.~Peris,
  {{Hyperasymptotics and quark-hadron duality violations in QCD}},
  \href{https://doi.org/10.1103/PhysRevD.97.054007}{{Phys. Rev. D}
  {\bf 97}, 054007 (2018)}, \href{https://arxiv.org/abs/1711.10316}{arXiv:1711.10316}

\bibitem{Beneke:2008ad}
M.~Beneke and M.~Jamin, $\alpha_s$ and the $\tau$ hadronic width:
  fixed-order, contour-improved and higher-order perturbation theory,
  \href{https://doi.org/10.1088/1126-6708/2008/09/044}{{JHEP} {\bf
  09}, 044  (2008)}, \href{https://arxiv.org/abs/0806.3156}{arXiv:0806.3156}


\bibitem{Baikov:2008jh}
P.~A. Baikov, K.~G. Chetyrkin and J.~H. K\"uhn, Order $\alpha^4_s$ QCD
  corrections to $Z$ and $\tau$ decays, \href{https://doi.org/10.1103/PhysRevLett.101.012002}{Phys. Rev. Lett.
  {\bf 101},  012002  (2008)}, \href{https://arxiv.org/abs/0801.1821}{arXiv:0801.1821}

\bibitem{Boito:2018rwt}
  D.~Boito, P.~Masjuan, and F.~Oliani, Higher-order QCD corrections to hadronic $\tau$ decays from Pad\'e approximants, \href{https://link.springer.com/article/10.1007%2FJHEP08%282018%29075}{JHEP {\bf 08}, 075  (2018)},
 \href{https://arxiv.org/abs/1807.01567}{arXiv:1807.01567} 

\bibitem{Jamin:2021qxb}
M.~Jamin, Higher-order behaviour of two-point current correlators, issue on Renormalons and Hyperasymptotics in QCD, EPJST, \href{https://arxiv.org/abs/2106.01614}{arXiv:2106.01614}

\bibitem{Bali:2013pla}
G.~S.~Bali, C.~Bauer, A.~Pineda and C.~Torrero, Perturbative expansion of the energy of static sources at large orders in four-dimensional SU(3) gauge theory,
\href{https://journals.aps.org/prd/abstract/10.1103/PhysRevD.87.094517}{Phys. Rev. D \textbf{87} (2013), 094517}, 
\href{https://arxiv.org/abs/1303.3279}{arXiv:1303.3279}

\bibitem{Bali:2013qla}
G.~S.~Bali, C.~Bauer and A.~Pineda, The static quark self-energy at $O(\alpha^20)$ in perturbation theory,
\href{https://pos.sissa.it/187/371}{PoS LATTICE 2013, 371 (2014)}, 
\href{https://arxiv.org/abs/1311.0114}{arXiv:1311.0114}

\bibitem{Ayala:2019hkn}
C.~Ayala, X.~Lobregat and A.~Pineda, Hyperasymptotic approximation to the top, bottom and charm pole mass,
\href{ https://journals.aps.org/prd/abstract/10.1103/PhysRevD.101.034002}{Phys. Rev. D \textbf{101}, 034002  (2020)},  \href{https://arxiv.org/abs/1909.01370} {arXiv:1909.01370}

\bibitem{Maiezza:2021mry}
A.~Maiezza and J.~Carlos Vasquez, Resurgence of the QCD Adler function, \href{https://www.sciencedirect.com/science/article/pii/S0370269321002781?via%3Dihub}{Phys. Lett. B \textbf{817}, 136338 (2021)}, 
\href{https://arxiv.org/abs/2104.03095}{arXiv:2104.03095}


\end{thebibliography}
\end{document}